\documentclass[prl,twocolumn,showpacs,preprintnumbers,amsmath,amssymb,superscriptaddress]{revtex4-1}

\usepackage{amsmath,amssymb,amsfonts} 	
\usepackage{graphicx}
\usepackage{hhline}
\usepackage[latin1]{inputenc}
\usepackage{subfigure}

\begin{document}

\title{Simple Screened Hydrogen Model of Excitons in Two-Dimensional Materials}

\author{Thomas Olsen}
\email{tolsen@fysik.dtu.dk}
\affiliation{Center for Atomic-Scale Materials Design and Center for Nanostructured Graphene (CNG), Department of Physics, Technical University of Denmark}

\author{Simone Latini}
\affiliation{Center for Atomic-Scale Materials Designg and Center for Nanostructured Graphene (CNG), Department of Physics, Technical University of Denmark}

\author{Filip Rasmussen}
\affiliation{Center for Atomic-Scale Materials Design and Center for Nanostructured Graphene (CNG), Department of Physics, Technical University of Denmark}

\author{Kristian S. Thygesen}
\affiliation{Center for Atomic-Scale Materials Design and Center for Nanostructured Graphene (CNG), Department of Physics, Technical University of Denmark}

\begin{abstract}
We present a generalized hydrogen model for the binding energies ($E_B$) of excitons in two-dimensional (2D) materials that sheds light on the fundamental differences between excitons in two and three dimensions. In contrast to the well-known hydrogen model of three-dimensional (3D) excitons, the description of 2D excitons is complicated by the fact that the screening cannot be assumed to be local. We show that one can consistently define an effective 2D dielectric constant by averaging the screening over the extend of the exciton. For an ideal 2D semiconductor this leads to a simple expression for $E_B$ that only depends on the excitonic mass and the 2D polarizability $\alpha$. The model is shown to produce accurate results for 51 transition metal dichalcogenides. Remarkably, over a wide range of polarizabilities the expression becomes independent of the mass and we obtain $E_B^{2D}\approx3/(4\pi\alpha)$, which explains the recently observed linear scaling of exciton binding energies with band gap. It is also shown that the model accurately reproduces the non-hydrogenic Rydberg series in WS$_2$ and can account for screening from the environment.
\end{abstract}
\maketitle

A striking property of two-dimensional semiconductors is the ability to form strongly bound excitons. This was initially predicted theoretically for hBN \cite{Wirtz2006}, graphane \cite{Cudazzo2010} and various transition metal dichalcogenides \cite{Huser2013, Komsa2012, Qiu2013} and has subsequently been confirmed experimentally \cite{Ye2014, Ugeda2014,hanbicki2015}. The quantum confinement of excitons in 2D comprises a tempting and intuitively appealing explanation for the large binding energies in these materials \cite{Yang1991a}. However, it is now well understood that the confinement of the local electronic environment in 2D plays a crucial role in the formation of strongly bound excitons \cite{Cudazzo2011, Huser2013}. The 2D electronic system is rather poor at screening interactions and the effective Coulomb interaction between an electron and a hole is simply much stronger in 2D than in 3D.

From a first principles point of view, the treatment of excitons requires advanced computational methodology such as the Bethe-Salpeter equation \cite{Albrecht1998, Rohlfing1998}. This approach has been applied to obtain absorption spectra for numerous insulators and usually yields very good agreement with experiments \cite{Onida2002}. However, only systems of modest size can be treated this way and simplified models of excitons will be an inevitable ingredient in calculations of realistic systems. For example, if the effect of substrates or dielectric environment is to be included in the calculation of excitons in 2D systems \cite{Andersen2015}, the computations become intractable with a standard Bethe-Salpeter approach. For three-dimensional materials the Mott-Wannier model comprises a strong conceptual and intuitive picture that provides a simple framework for calculating exciton binding energies \cite{Wannier1937}. In the center of mass frame, an excited electron-hole pair can be shown to satisfy a hydrogenic Schr{\"o}dinger equation, where band structure effects are included through an excitonic effective mass $\mu$ and the dielectric screening from the environment is included through the static dielectric constant $\epsilon_0$. The exciton binding energy in atomic units is then written as
\begin{align}\label{eq:e_3d}
E_B^{3D} = \frac{\mu}{2\epsilon_0^2}.
\end{align}
Thus the daunting task of solving the Bethe-Salpeter equation, has been reduced to the calculation of just two parameters: the effective mass and the static dielectric constant, both of which are easily obtained with any standard electronic structure software package. This approximation is well justified whenever the screening is local such that its Fourier transform can be approximated by a constant in the vicinity of the origin. However in highly anisotropic structures such as layered materials this assumption is expected to break down.

In 2D dielectrics, it is well known that the screening takes the form $\epsilon(\mathbf{q})=1+2\pi\alpha q$ \cite{Cudazzo2010}, where $\alpha$ is the 2D polarizability. The screening is thus inherently non-local and it is not obvious if it is possible to arrive at a hydrogenic model like Eq. \eqref{eq:e_3d}. Instead one can calculate the 2D screened potential and solve the Schr{\"o}dinger equation for the electron-hole wavefunction: 
\begin{align}\label{eq:schrodinger}
\Big[-\frac{\nabla^2}{2\mu}+W(\mathbf{r})\Big]\psi(\mathbf{r})=E_n\psi(\mathbf{r}),
\end{align}
where $W(\mathbf{r})$ is the 2D convolution of the Coulomb interaction and $\epsilon^{-1}(\mathbf{r}-\mathbf{r}')$. However, in general this may be a tedious task and it would be highly desirable to have an expression like Eq. \eqref{eq:e_3d} from which the exciton binding energy in a given material can be easily estimated and understood. To accomplish this, we calculate the average screening felt by the exciton. To this end we consider the expression
\begin{align}\label{eq:eps_int}
\epsilon_{eff} = \frac{a_{eff}^2}{\pi}\int_0^{2\pi}d\theta\int_0^{1/a_{eff}}dq q\epsilon(\mathbf{q}),
\end{align}
where $a_{eff}$ is the effective Bohr radius. For the 2D hydrogen atom the Bohr radius is given by $a=\epsilon/(2\mu)$ and Eq. \eqref{eq:eps} has to be solved self-consistently for $\epsilon_{eff}$ given an expression for $\epsilon(\mathbf{q})$. In a strictly 2D system the screening is linear in $q$ and Eq. \eqref{eq:eps_int} can be solved to yield
\begin{align}\label{eq:eps}
\epsilon_{eff} = \frac{1}{2}\Big(1+\sqrt{1+32\pi\alpha\mu/3}\Big).
\end{align}
Using that the hydrogenic binding energy in 2D is a factor of four larger than in 3D \cite{Yang1991a}, we obtain
\begin{align}\label{eq:e_2d}
E_B^{2D} = \frac{8\mu}{\big(1+\sqrt{1+32\pi\alpha\mu/3}\big)^2}.
\end{align}
This is the main result of the present letter and comprises a long-sought-for 2D analog of Eq. \eqref{eq:e_3d}.

\begin{figure}[tb]
    \includegraphics[width=7.0 cm]{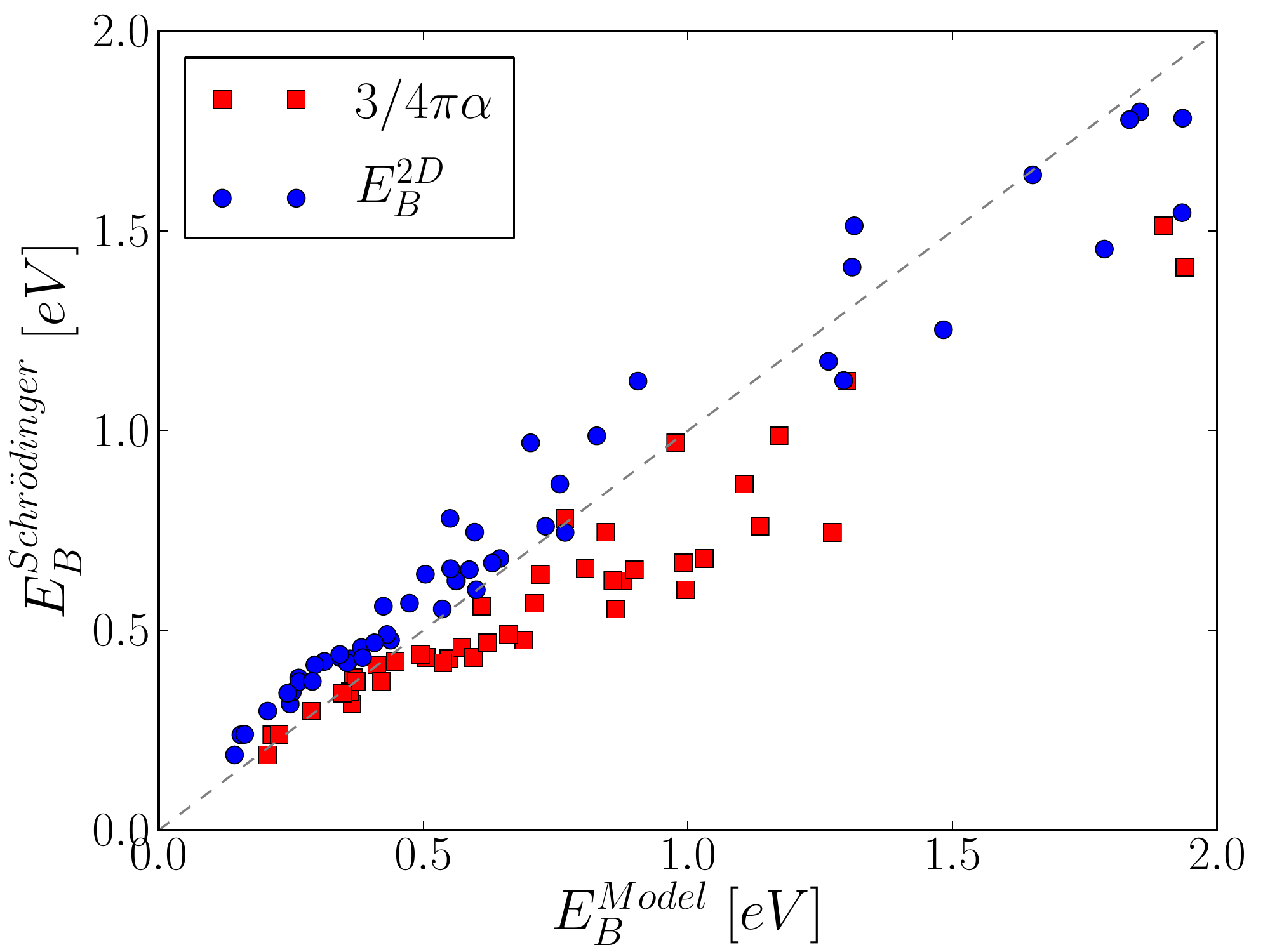}
\caption{(Color online) Exciton binding energies of 51 transition metal dichalcogenides calculated as the lowest eigenvalue of Eq. \eqref{eq:schrodinger} (vertical axis) and the model result Eq. \eqref{eq:e_2d} (horizontal axis).}
\label{fig:eb_eb}
\end{figure}
A remarkable property of the expression \eqref{eq:e_2d} is the fact that it becomes independent of the effective mass if the polarizability is large. More precisely
\begin{align}\label{eq:e_2d_app}
E_B^{2D}\approx\frac{3}{4\pi\alpha},\qquad32\pi\alpha\mu/3\gg1.
\end{align}
It may come as a surprise that the binding energy becomes independent of mass, since a large mass gives rise to a localized exciton and the binding energy typically increases with localization. This is reflected in Eq.  \eqref{eq:e_3d}, where the binding energy is seen to be proportional to the mass. However, in 2D, short range interactions are screened more effectively than long range interactions. Thus, there are two opposing effects of the exciton mass and for large polarizabilities, the binding energy becomes independent of mass. In order to assert the applicability of the expressions \eqref{eq:e_2d}-\eqref{eq:e_2d_app}, we have calculated the static polarizability and effective masses of 51 semiconducting monolayers of transition metal dichalcogenides within the Random Phase Approximation. The calculations were performed with the electronic structure code GPAW \cite{gpaw-paper, Yan2011}, which is based on the projector augmented wave formalism. Further details on the calculations can be found in Ref. \cite{Rasmussen2015}. In Fig. \ref{fig:eb_eb} we compare the model binding energies with the full solution of Eq. \eqref{eq:schrodinger}. Using the expression \eqref{eq:e_2d}, the agreement is seen to be on the order of 10{\%}. For materials with anisotropic mass tensor we have used the average mass, both in the model and when solving the Schrodinger equation. With the approximated expression \eqref{eq:e_2d_app}, we obtain excellent agreement for binding energies up to $\sim0.5$ eV, whereas the binding energies are underestimated for strongly bound excitons.

\begin{figure}[tb]
    \includegraphics[width=7.0 cm]{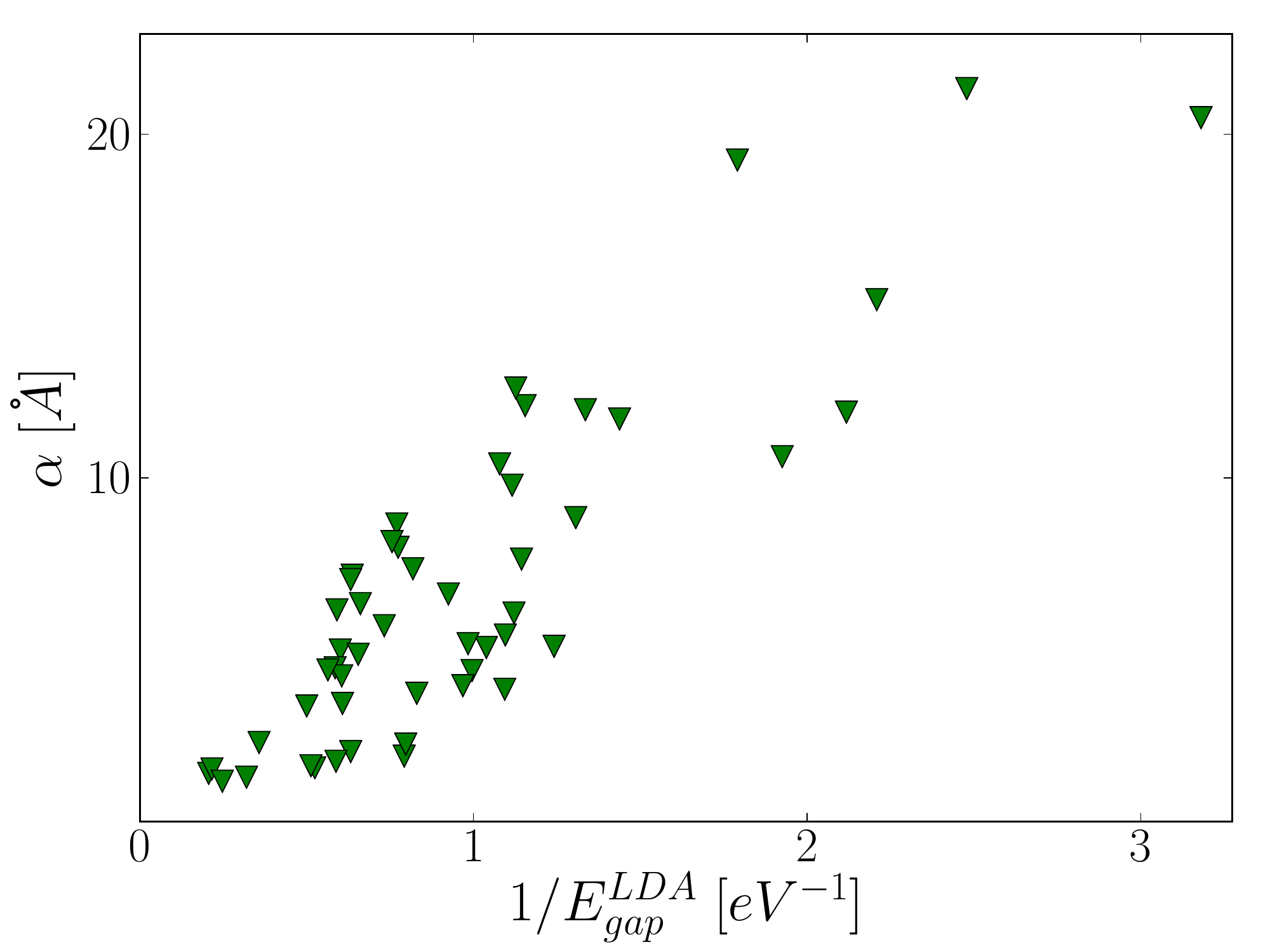}
\caption{(Color online) The 2D polarizability of 51 transition metal dichalcogenides shown as a function of LDA band gaps.}
\label{fig:alpha}
\end{figure}

Recently, first principles calculations have indicated that exciton binding energies in different 2D materials scale linearly with the band gaps \cite{Choi2015}. In the present model, this behavior comes out naturally since without local field effects, the in-plane components of the polarizability in the Random Phase Approximation are given by
\begin{align}\label{eq:alpha}
\alpha=\sum_{m,n}\int_{BZ}\frac{d\mathbf{k}}{(2\pi)^2}(f_{n\mathbf{k}}-f_{m\mathbf{k}})\frac{|\langle u_{m\mathbf{k}}|\hat r_\parallel|u_{n\mathbf{k}}\rangle|^2}{\varepsilon_{n\mathbf{k}}-\varepsilon_{m\mathbf{k}}},
\end{align}
and we expect that $\alpha$ will be roughly inversely proportional to the band gap. This is illustrated in Fig. \ref{fig:alpha} for the 51 transition metal dichalcogenides. Combining this with Eq. \eqref{eq:e_2d_app} thus gives $E_B^{2D}\propto E_{gap}$. However, in the present model the scaling originates solely from the screening and not the effective mass as previously proposed \cite{Choi2015}. For the present set of materials, we do not observe any correlation between binding energies and effective mass. We use the LDA band gaps and not the quasiparticle gaps, which could be obtained from for example GW calculations \cite{Rasmussen2015}, since LDA typically gives a better estimate of the two-particle excitation energies that enters the expression for $\alpha$. In contrast, the use of GW gaps would underestimate the screening due to the lack of electron-hole interactions.

\begin{figure}[tb]
    \includegraphics[width=7.0 cm]{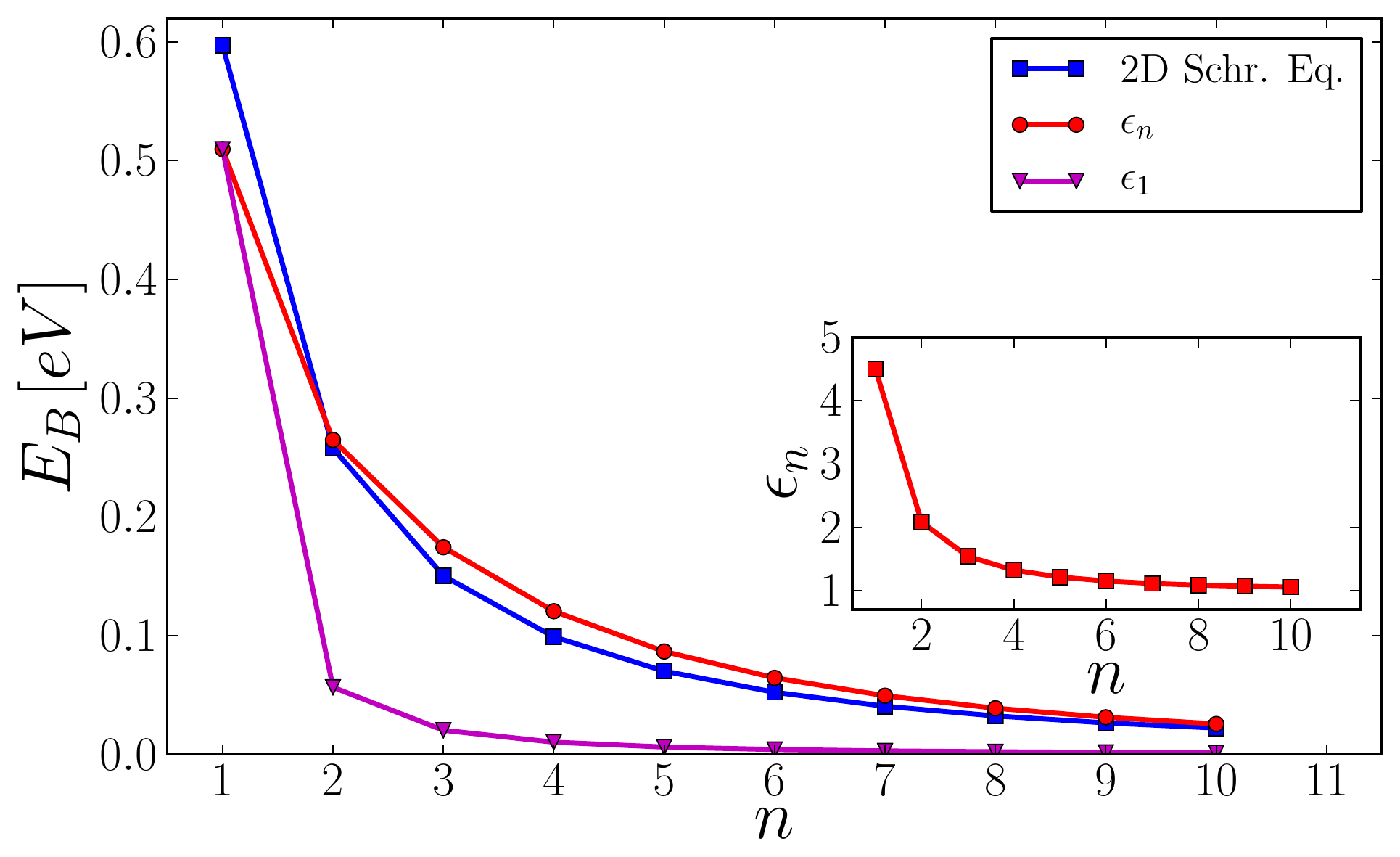}
\caption{(Color online) Rydberg series of a monolayer of 2H-WS$_2$ calculated with the generalized hydrogen model with linear screening (Eqs. \eqref{eq:E_n} and \eqref{eq:eps_n}) and from the solution of the 2D screened Schr{\"o}dinger equation \eqref{eq:schrodinger}. The results are compared with the bare hydrogen model where the effective screening obtained from the ground states is used for all states.}
\label{fig:rydberg}
\end{figure}

To validate the general applicability of the effective screening model, we now show that it can also be used to account for the Rydberg series in 2D materials. In Ref. \cite{Chernikov2014}, the Rydberg series in WS$_2$ was measured and shown to deviate significantly from the Rydberg series of a 2D hydrogen model scaled by an overall screening factor. The reason is simply that the effective screening depends on the $n$ quantum number due to the increasing spatial extend of higher lying Rydberg states. The authors used the results to define an $n$-dependent effective screenings $\epsilon_n$, which were then determined by fitting each term in the Rydberg series to a 2D hydrogen model. The Rydberg series is then given by
\begin{align}\label{eq:E_n}
E_n^{2D}=-\frac{\mu}{2(n-\frac{1}{2})^2\epsilon_n}.
\end{align}
Two of the present authors have recently showed that the Rydberg series can accurately be reproduced by solving Eq. \eqref{eq:schrodinger} with a screened 2D potential calculated from first principles \cite{Andersen2015} and we will assume that approach to be an accurate reference. Here we calculate the $n$-dependent effective screening from first principles by replacing $a_{eff}$ in Eq. \ref{eq:eps_int} by an $n$-dependent characteristic extension of the state. To this end we note that for $l=0$, the first moment of a state with principal quantum number $n$ in a 2D hydrogen atom is \cite{Yang1991a}
\begin{align}
a_n\equiv\langle n|\hat r|n\rangle=[3n(n-1)+1]/(2\mu),
\end{align}
where $\hat r=\sqrt{\hat x^2 +\hat y^2 }$. In terms of this, the $a_{eff}$ defined previously is given by $a_1$ and $E_B^{2D}$ is $-E_1^{2D}$. Within the linear model the effective screening for state $n$ then becomes
\begin{align}\label{eq:eps_n}
\epsilon_n = \frac{1}{2}\bigg(1+\sqrt{1+\frac{32\pi\alpha\mu}{9n(n-1)+3}}\bigg).
\end{align}
It is straightforward to generalize these expressions to $l\neq0$ \cite{Yang1991a}, which results in a larger value of the effective radius $a_{nl}$ and thus $\epsilon_{n,l>0}<\epsilon_{n,l=0}$. The energy is still given by Eq. \eqref{eq:E_n} and at a given $n$, the higher angular momentum excitons will therefore have a larger binding energy, which has been observed in the case of 2H-WS$_2$ monolayers \cite{Ye2014}. As a case study we consider this material and apply the linear screening model. We obtain a first principles 2D polarizability of $\alpha=5.25$ {\AA} and $\mu=0.19$. In Fig. \ref{fig:rydberg} we show the Rydberg series calculated with the generalized hydrogen model, which agrees very well with a full solution of Eq. \eqref{eq:schrodinger}. In contrast, the pure 2D hydrogen model with an overall effective screening is seen to significantly underestimate the binding energies at higher lying states, since the decreased screening of extended states is not taken into account. We also note that the model binding energies of the $n=1$ state agree very well with a full solution of the Bethe-Salpeter equation which yields an exciton binding energy of 0.54 eV \cite{Shi2013}.

\begin{figure}[t]
    \includegraphics[width=7.0 cm]{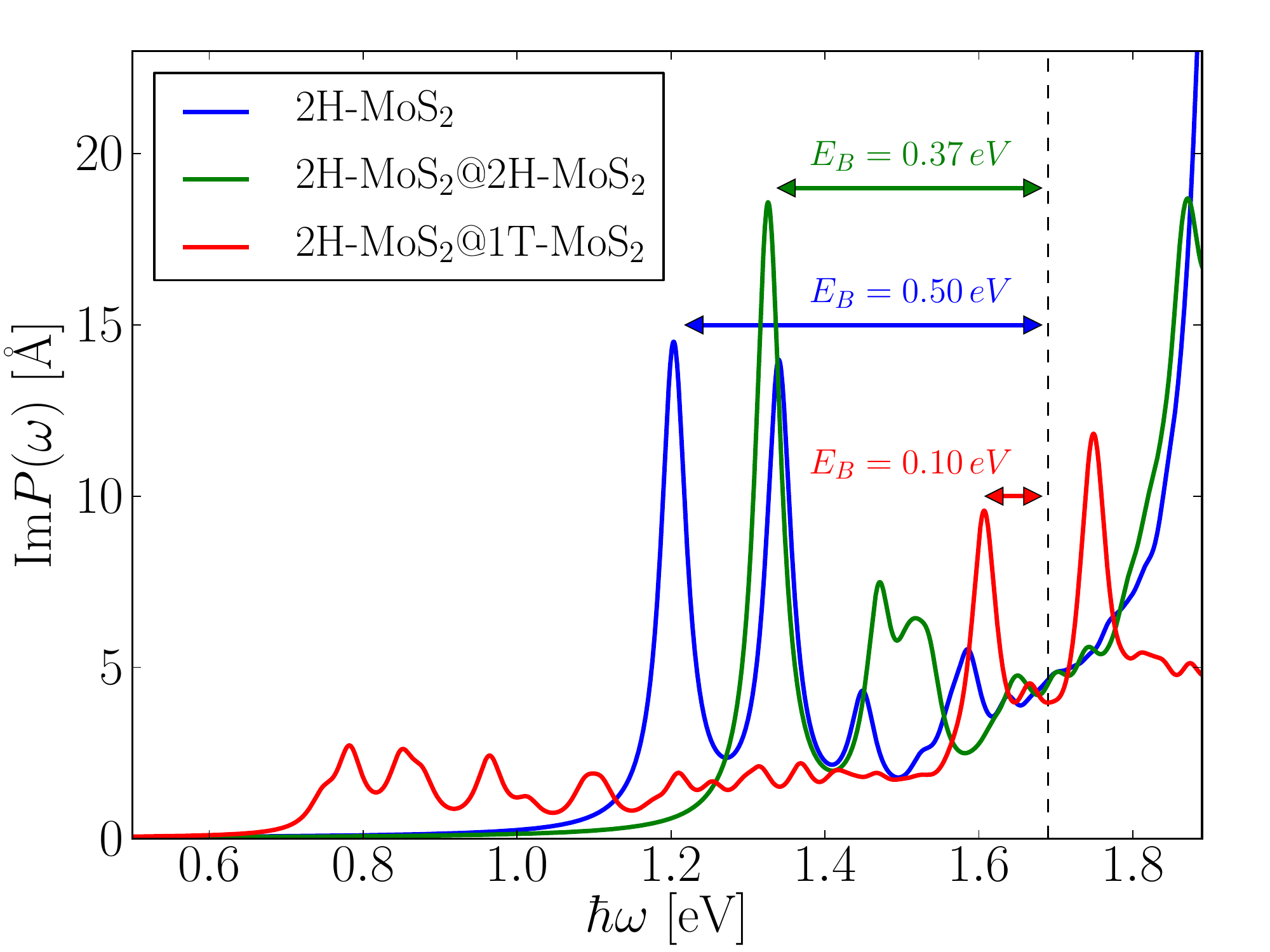}
\caption{(Color online) Dynamic 2D polarizability of 2H-MoS$_2$ in different environments calculated from the Bethe-Salpeter equation based on Kohn-Sham eigenvalues. The vertical lines at 1.7 eV marks the Kohn-Sham band gaps, which are nearly identical in the three cases.}
\label{fig:bse}
\end{figure}

We now proceed to show how the effect of screening by the environment can naturally be taken into account in the present framework. It should be noted, however, that the linear model for the screening is expected to break down for systems where the vertical extend of a substrate becomes comparable to the Bohr radius of the exciton. For example, if we consider a stack of $N$ monolayers, $\alpha$ will diverge in the limit of large $N$, since the bulk system will have $\epsilon(\mathbf{q=0})\neq1$ \cite{Andersen2015, Latini}. The linear regime will therefore only be valid at infinitesimal values of $\mathbf{q}$ when $N$ becomes large. As an example where we expect the linear model to be applicable we consider a monolayer 2H-MoS$_2$ and compare the isolated layer with the two cases where it is in the vicinity of another layer of 2H-MoS$_2$ and in the vicinity of a metallic layer of 1T-MoS$_2$. In Fig \ref{fig:bse}, we show the absorption spectrum calculated from the Bethe-Salpeter equation based on Kohn-Sham eigenvalues. The BSE calculations were performed in a plane wave basis with a 2D Coulomb truncation scheme \cite{Rozzi2006, Huser2013a} using a $60\times60$ $k$-point mesh. It is well known that the low energy absorption spectrum of this system exhibits a double excitonic peak due a spin-orbit split valence band \cite{Mak2010, Molina-Sanchez2013}. This facilitates the identification of the excitons in the 2H-MoS$_2$ layer in the vicinity of a metallic substrate with low lying excitations. We have not performed the full spinorial BSE calculations, but simply included spinorbit effects in the band structure in order to identify the excitons. In the following we consider the binding energies of the lowest exciton. The isolated layer exhibits an exciton bound by 0.50 eV. In the vicinity of another 2H-MoS$_2$ layer, the binding energy is decreased to 0.37 eV and the metallic 1T-MoS$_2$ decreases the binding energy to 0.10 eV. We note that the quasiparticle band structure corrections are expected to be much smaller for the case of 2H-MoS$_2$@1T-MoS$_2$ such that the actual positions of the excitons would be similar for the three cases in an optical absorption experiment. However, we have chosen to leave out the quasiparticle corrections in order to illustrate the difference in binding energies more clearly. 

To apply the model we wish to calculate $\epsilon(\mathbf{q})$ for the 2H-MoS$_2$ layer when it is in the vicinity of a screening environment. For small $q$, we may still write it as $\epsilon(\mathbf{q})=1+2\pi\tilde\alpha q$ and we would like to extract $\tilde\alpha$, which is the relevant quantity for the screened hydrogen model. 
We calculate it by the finite difference
\begin{align}
2\pi\tilde\alpha=\epsilon(\mathbf{q}_1)-1,
\end{align}
where $\mathbf{q}_1$ is a small finite value of $q$. In the present case we take $\mathbf{q}_1$ as the smallest $q$-vector in the direction of $K$ obtained from a $60\times60$ $k$-point grid. The 2D dielectric function is obtained from 
\begin{align}\label{eq:eps_av}
\frac{1}{\epsilon(\mathbf{q})}=\frac{\langle V_{tot}(\mathbf{r_\parallel},z_0)e^{-i\mathbf{q}\cdot\mathbf{r}}\rangle_A}{V_\mathbf{q}},
\end{align}
where $V_{tot}(\mathbf{r})$ is the total potential resulting from an external perturbation $V_{ext}(\mathbf{r})=V_\mathbf{q}e^{i\mathbf{q}\cdot\mathbf{r}}$ and $\langle\ldots\rangle_A$ denotes average over the 2D unit cell of area $A$. It is straightforward to relate this expression to an average over the microscopic dielectric function $\epsilon^{-1}(\mathbf{r},\mathbf{r}')$, which can be calculated in the Random Phase Approximation by most electronic structure codes. We take $z_0$ to be at the center of the 2H-MoS$_2$ layer, but we note that $\tilde\alpha$ is approximately independent of the value of $z_0$ when $z_0$ chosen in any part of the central 3.0 {\AA} of the layer. In Tab. \ref{tab}, we display the calculated values of $\tilde\alpha$ along with the exciton binding energies obtained from the model \eqref{eq:e_2d}, the 2D Schr{\"o}dinger equation \eqref{eq:schrodinger}, and the BSE calculations. As expected, the environment strongly affects the value of $\tilde\alpha$. In particular, the metallic 1T-MoS$_2$ layer significantly increases the screening, whereas the presence of another 2H-MoS$_2$ layer results in a less pronounced effect. We find good agreement between the simple model, the 2D Schr{\"o}dinger, and the BSE calculations. We should note that the convergence of the exciton binding energies in the presence of the metallic 1T-MoS$_2$ layer is very slow with respect to $k$-point sampling and the converged result is expected to exhibit a lower binding energy than the one obtained here. Furthermore, we have not included the intraband contribution to the static screening, which, is expected to scale as $\sim1/q$ in 2D metals. In fact it is not clear that the dynamic contributions to the screening can be neglected in the either the BSE approach or the model. On the other hand, the 1T structure is known to distort into the so-called 1T' structure, which is a topological insulator with a gap on the order $50$ meV \cite{Qian2014}. In any case, the screening is treated at the same footing in the BSE and the model calculations since the values of $\tilde\alpha$ were obtained by a finite difference calculation on the same $k$-point grid that was used in the BSE calculations. Nevertheless, the model is easily generalized to a non-linear $\epsilon(\mathbf{q})$, the only difference being that \eqref{eq:eps_int} should be solved numerically. We note again that the applicability of this approach is limited to cases where the linear model is expected to be agood approximation. For extended substrates, the present approach may be generalized by calculating the full $\epsilon(\mathbf{q})$ and solving Eq. \eqref{eq:eps_int} numerically, but it is not clear that the analytical results derived from the 2D hydrogen model \eqref{eq:E_n} is able to produce reliable results in this case. Alternatively one may solve a quasi-2D Schr{\"o}dinger equation that incorporates the finite extend of the slab \cite{Latini}.

\begin{table}[tb]
\begin{center}
\begin{tabular}{c|c|c|c}
      & 2H-MoS$_2$ & 2H-MoS$_2$@2H-MoS$_2$ & 2H-MoS$_2$@1T-MoS$_2$ \\ 
      \hhline{=|=|=|=}
$E_B^{BSE}$ [eV]     & 0.50     & 0.37   & 0.10  \\
      \hhline{=|=|=|=}
$E_B^{Schr.}$ [eV]   & 0.58     & 0.40   & 0.17  \\
      \hline
$E_B^{Model}$ [eV]   & 0.48     & 0.30   & 0.10  \\
      \hline
$\tilde\alpha$ [{\AA}]  & 5.83   & 10.0    & 30.1
\end{tabular}
\caption{Exciton binding energies for 2H-MoS$_2$ in different environments calculated from the Bethe-Salpeter equation (BSE), the 2D Schr{\"o}dinger equation, and the generalized screened hydrogen model. We also display the values of $\tilde\alpha$, which is the polarizability of the single 2H-MoS$_2$ layer used in the calculations. For all calculations we used an effective exciton mass of 0.276, which was obtained from the \text{ab initio} band structure.}
\label{tab}
\end{center}
\end{table}
To conclude, we have presented an analytical expression for the exciton binding energies in 2D semiconductors that only depends on the static 2D polarizability and the effective mass and produces quantitative agreement with the solution of the full screened 2D Schr{\"o}dinger equation. It has also been shown that for large polarizabilities, the result becomes independent of mass and yields a linear relation between exciton binding energies and band gaps. It has previously been anticipated that the non-hydrogenic Rydberg series could be attributed to an $n$-dependent value of the effective screening \cite{Chernikov2014}. Here we have obtained an explicit expression for $\epsilon_n$ that provides an accurate account of the full exciton spectrum. It has also been shown, that the model can be generalized to incorporate the effect of a simple screening environment. We do not claim that the presented expression for the effective screening \eqref{eq:eps_int} in the linear model is unique. In fact, it is based on an unweighted average of a linear model for the non-local 2D screening over the extend of the exciton and it is easy to imagine more elaborate averaging schemes. However, we believe that the simplicity is the main merit of this procedure and the resulting analytical expressions are very easy to apply to a given 2D material. In particular, for complicated structures it may not be possible to treat the electron-hole interaction by a first principles approach and our model results could be a crucial ingredient in understanding the excitonic structure in such materials.

The authors acknowledge support from the Danish Council for Independent Research's Sapere Aude Program through grant no. 11-1051390. The Center for Nanostructured Graphene (CNG) is sponsored by the Danish National Research Foundation, Project DNRF58.


%

\end{document}